# High Sensitive α-Fe$_2$O$_3$ Nano-structured Gas Sensor Fabricated through Annealing Technique for Detecting Ethanol

*Hamed Aleebrahim Dehkordi [a], Ali Mokhtari*[a, b], Vishtasb Soleimanian[a, b] and Mohsen Ghasemi[a, b]*

*[a] Department of Physics, Faculty of Sciences, Shahrekord University, P.O. Box 115, Shahrekord, Iran*

*[b] Nanotechnology Research Institute, Shahrekord University, 8818634141 Shahrekord, Iran*

Abstract

On the way to advance the sensing technology, various strategies based on the nano-materials have been introduced to improve the performance of the gas sensors. In this study, we have introduced a facile fabrication procedure for annealing hematite to monitor ethanol gas. Due to large specific area of this conductive platform, more available target molecules (ethanol gas) are detected. To construct this platform, initially, a large scale well-separated iron oxide (hematite) nanostructures are created via an annealing process. The morphology of the nanostructure is optimized through annealing temperature and operating temperature for ethanol gas. The best response for the gas sensor is achieved for hematite nanostructures fabricated through the annealing temperature of 500°C and the operating temperature of 225°C. The fabricated nanostructures is tested in air ambient conditions for various ethanol gas concentrations from 50 to 1000 ppm. The modified sensor exhibited acceptable reproducibility and good selectivity with no interference-effect. To examine the X-ray diffraction patterns of hematite nanostructures at three annealed temperatures, the Rietveld method and FullProf software are used in which the average crystal size has been obtained. Obviously, with increasing the annealing temperature, the average size of the crystals has been increased. The results of ultraviolet-visible spectroscopy show that the energy gap decreased with increasing annealing temperature. The maximum sensing response of hematite flower-like nanostructures is obtained at a concentration of 1000 ppm. Also, the shortest response time is estimated at about 27 seconds for ethanol at a concentration of 200 ppm. The detection limit of this sensor is obtained at 50 ppm.

1. Introduction

Ethanol has played an important role in the development of human civilization and it is intensively used in our daily lives from biomedical to the automotive fuel industry [1]. However, releasing large quantities of this flammable and toxic gas is resulted in polluting ecosystems. Since the human health has been influenced by breathing high concentrations of ethanol [2], a precise monitoring of the exposure level of ethanol gas is essential to control the public safety and also the environmental protection. Currently, considerable efforts have been devoted for developing the gas sensor characteristics including low cost, rapid response, high sensitivity, simplicity, good time recovery, and selectivity.
The response of these gas sensors is attributed to an intelligent selection of used material and the surface designing procedure which have played a critical role for detecting an extremely low trace level of ethanol. Up to now, a large number of metal oxide including ZnO, TiO$_2$, CuO, and SnO$_2$ have been used in gas sensors [3-10]. However, the disadvantages of these

---

metal oxides are their low sensitivity and poor selectivity between present gases in detecting sample gas. Among these metal oxides, hematite (α-$Fe_2O_3$) as a nontoxicity, highly stable, a multifunctional n-type semiconductor with a band gap of 2.1eV has received extensive attention owing to its good interstice physical and chemical properties [11-13]. It has been widely investigated in many practical fields including magnetic devices, photo-catalyst fields, gas sensing, photo- anode, super-capacitor and rechargeable lithium-ion batteries [1, 14-20].

In order to improve the chemical and physical properties of the hematite, many studies have been focused on designing various hematite morphology in different sizes and shapes to enhance the gas sensor performance. The superior performances of porous metal oxide compared to their solids have been studied in many articles in various fields such as catalysts, biomedical and chemical sensors. It is worthwhile to mention that the nanoscale structures of the α- $Fe_2O_3$ have been represented strongly better performance [21]. The surface-to-volume ratio and effective surface area of the porous structure are increased at nanoscale leading to the enhancement of material surface reaction efficiency. Therefore, the porous metal oxides are considered as more convenient and effective candidates for sensor materials in comparison with solid ones [1, 14, 16-18, 22-23]. Up to now, a large number of techniques have been used to synthesis various structures such as nanoparticles [24-26], nanorods [27], nanowires [20, 28], nano-sheets [28-30], nano-spheres [31-32], nano-plates [33-34], nanoflowers [35-36], nano-crystals [37-38], polyhedral nanoparticles [39], nano-ribbons [40], nano-tubes [41-42], nanostructured microspheres [42-43] and hollow nanostructures [31].

Besides, to fabricate the α- $Fe_2O_3$ nanostructures, a lot of conventional methods have been practiced including electrochemical deposition, vapor solid liquid (VLS), sol gel, template method, ionic layer adsorption and annealing approach [44-53]. Among these methods, hydrothermal synthesis and the annealing process are ideal methods due to the creation of homogeneous metal structures on the solid support. During the annealing process, at the temperature close to the transition temperature of the solid substrate, both surface diffusion, and recrystallization rate are formed. Therefore, continuous morphology is transported into the ensembles of individual nuclei with sub micro-scale lengths which are embedded on the solid support. The surface morphologies can be optimized by controlling annealing temperature.

In the present work, we have studied the role of annealing temperature on the morphology of hematite. The fabricate mechanism has been done through a two-step approach. First, with an application of ferric chloride solution ($FeCl_3.6H_2O$) and sodium nitrate ($NaNO_3$) solution, the FeO-OH layer has been fabricated on the glassy substrate. Then as a second step, the annealing process where hematite with porous morphology has been constructed. This best fabricated nanostructure obtained through 500°C has been utilized for detecting ethanol with various operating temperatures. The best response of our gas sensor has been achieved at 225°C. The selectivity and fast-response of the α- $Fe_2O_3$ sensor toward ethanol have been also studied for various concentrations from 50 to 1000 ppm.

2. Experiment
   2.1.   Materials and instruments

The $FeCl_3$-$6H_2O$, $NaNO_3$, $C_2H_5OH$, HCl and acetone ($CH_3$)$_2$ CO are supplied from Merck products. A normal glass is used as a substrate to grow α-$Fe_2O_3$ nanostructure. All chemical materials are at an analytical grade. The aqueous solutions are prepared using double deionized (DI) water with specific resistance of 18.2 MΩ at 25°C.

The surface morphological study of the various hematite films is characterized by a field emission scanning electron microscopy FESEM, FEI Nova Nano-SEM-450 model. The x-ray diffraction (XRD) patterns are obtained at room temperature with a Philips Xpert MPD diffractometer equipped with Cu-kα operated at 40keV and 30 mA in a scanning range of 2θ (3°-80°). The JCPDS Date File is used for recognizing the diffraction peaks of the obtained crystalline phase. The absorbance spectrum is studied using UV–Vis spectrophotometer (+T80, PG Instruments, England) in the range of 200–800 nm.

## 2.2. Synthesis of the α-Fe$_2$O$_3$ nanostructures

The surface modification of the α-Fe$_2$O$_3$ is constructed in two facile steps: hydrothermal synthesis of the FeO-OH and annealing treatment of the FeO-OH. Prior to synthesis, the glasses used as substrates are cut into 20 mm×15 mm pieces and are cleaned in the deionized water (DW), acetone and ethanol through an ultrasonic bath for 15 minutes, respectively. Then, the glasses are placed in an oven at the temperature of 80°C to be dried and then left to become cool to the room temperature.

To prepare the FeO-OH layer, an aqueous mixture containing 0.2 ml ferric chloride solution (FeCl$_3$-6H$_2$O) and 1 ml of sodium nitrate solution (NaNO$_3$) are prepared at stirring condition at the room temperature for about 30 minutes (1.08 g of chloride powder and 1.699 g of sodium nitrate powder are dissolved in 20 ml of the DW). In order to adjust the PH of the mixture to 1.5, sufficient amounts of hydrochloric acid are also added. The stirring condition has been chosen with the aim of homogenization.

Then, the clean glasses are immersed in a vertical position into a Teflon autoclave containing the above mixture and finally, the Teflon autoclave is placed in a furnace at the temperature of 95°C for 20 hours, and then left to become cool to the room temperature. The following reaction (equation 1) is described the hydrolysis of the aqueous Fe$^{3+}$ ions with OH$^-$ to produce iron oxide nuclei under hydrothermal conditions [54].

$$Fe^{3+}_{(aq)} + 3OH^-_{(aq)} \rightarrow \beta - FeOOH_{(s)} + H_2O \qquad (1)$$

After the hydrothermal reaction, a uniform yellow thin layer of the β-FeO-OH is observed on the glass substrate (akageneite-coated glass). The glass substrate is thoroughly rinsed with DW to remove any residual salts and then is dried at 60℃ for about half an hour.

In order to convert to hematite and obtain α-Fe$_2$O$_3$ nanostructures, the akageneite-coated substrates are placed in a furnace for annealing treatment in air at various temperature (400°C, 500°C and 600°C) for 2 hours.

It should be noted that after the annealing process, the substrates color changed to a reddish brown color, indicating the creation of a new FeO-OH phase due to lost water and its conversion to iron oxide (hematite). The reactions that take place in the annealing process are presented in as the following equation:

$$2\beta - FeOOH_{(s)} \rightarrow Fe_2O_{3(s)} + H_2O_{(l\ or\ g)} \qquad (2)$$

### 2.3. Gas sensor preparation and its test chamber

To evaluate the gas sensor properties and also to measure the sensitivity of the modified samples, the physical vapor deposition technique (PVD) has been used to fabricate the circuit board substrate. According to the designed mask, the gold pattern is fabricated on the surface of iron oxide nanostructures (hematite) nanostructures in the PVD system. The outline dimension of the circuit board substrate is about 10*15 mm$^2$.

The sensing tests used in this research are performed in dynamic mode by a gas sensor which the device details are recorded in our previous report [55]. Briefly, the gas sensor system consists of different components including gas and air inlet and outlet chambers, flow controller, electrode connections, valves, heater, thermometer, and ohmmeter which are connected to a computer. A heater and a digital thermometer are used for measuring the instantaneous reading temperature of the sensing layer. The connectors are connected to the metal electrodes for ohmic contact. The resistance has been continually read, and the data is transmitted to the computer for recording. The electrical measurements are recorded by two gold electrodes as contacts placed face to face on opposite sides of hematite film through the thermal evaporation technique (PVD). The sensing measurements for all nanostructure based on the $Fe_2O_3$ films are carried out in fixed lab environmental conditions of 27°C and relative humidity of about 40%.
To obtain a stabilized resistance, the hematite films are kept in the test chamber at the operating temperature for 20 minutes before each test. Hence, $R_a$ values are derived. Afterward, gas is pumped into the sealed chamber and $R_g$ values are calculated dynamically [55]. The $R_a$ and $R_g$ are resistance values of the thin film in air and in presence of gas, respectively.

### 3. Results and discussion

#### 3.1. Morphological and structural characterization of the gas sensor surface

The morphology of the synthesized iron oxide thin films is characterized using FESEM images. In figure 1a, the morphologies of the substrate before the annealing process are shown. The surface morphology of the annealed $Fe_2O_3$ at different temperatures from 400°C to 600°C is illustrated in figure 1b-1d. It is clear that after the annealing process, the surface morphology changed dramatically. The FESEM images showed that the temperature has significantly affected the formation of hematite flower-like nanostructures and their growth rate. The surface morphology of iron oxide at an annealing temperature of 400°C is shown in figure 1b. At this temperature, small granules with a thickness of about 100 nm are densely packed together. By increasing the annealing temperature from 400°C to 500°C, the surface morphology of the deposited iron oxide is converted to flower-like morphology by increasing the grain length. The morphology of the flower species has a needle head, which increased the effective surface sample. This special morphology with more effective site led to an efficient sensing process. As the annealing temperature is varied from 500°C to 600°C, the surface structure of the 600°C annealed sample is so dense and the thickness of the flower leaf is also increased so that the oblique and irregular columns are formed. It is worthwhile to mention that no needle head is observed on the flower leaves at this temperature.

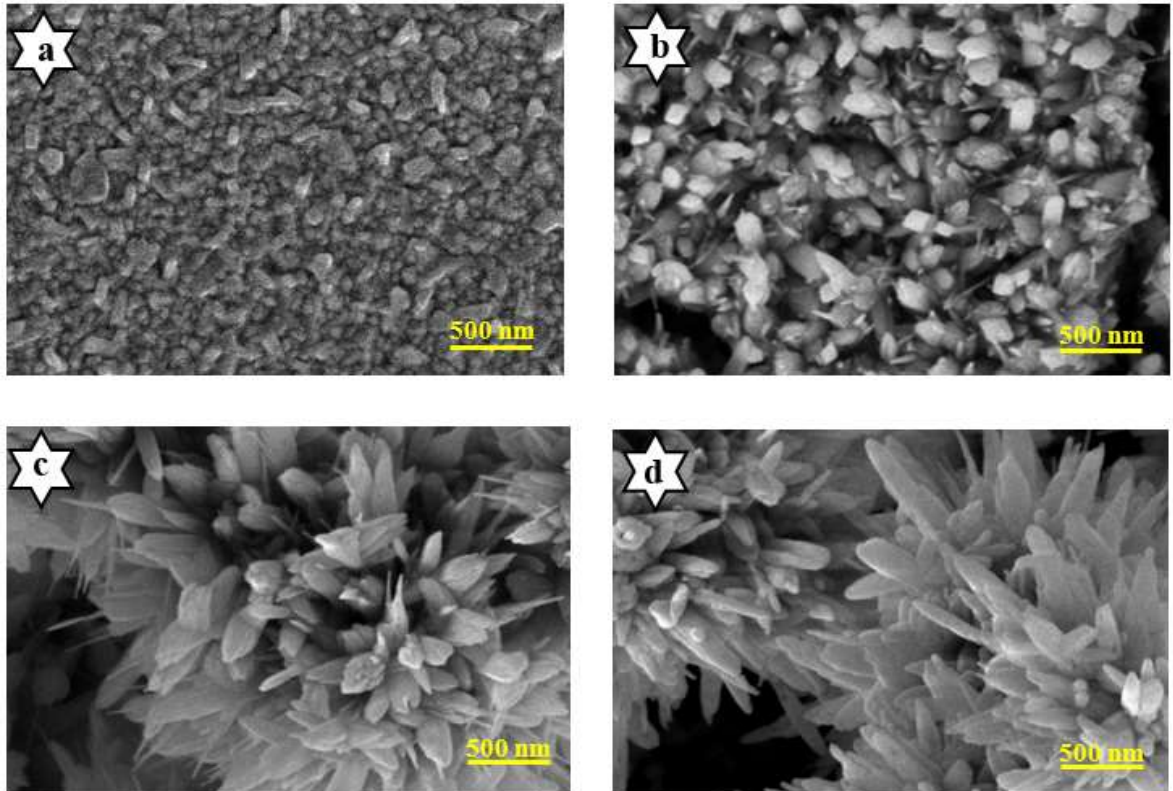

Figure 1: The top view of FESEM images of a) the sample before the annealing process, b) 400°C, c) 500°C, d) 600°C annealed $Fe_2O_3$. The scale bar of the FESEM images is set to be 500 nm.

The formation of iron oxide nanostructures on the glass substrate is supported by the XRD patterns. In figure 2, the crystallographic orientation of different annealing temperatures at 400°C, 500°C and 600°C on the glass substrate are shown. The pattern of iron oxide (hematite) at three annealed temperatures is fully consistent with the JCPDS standard card number 0664-33. Iron oxide (hematite) belongs to the spatial group of R-3c with lattice parameters a=5.036, b =5.036 and c =13.74 $A^0$. The X-ray diffraction patterns displayed the crystal orientation enhancement in the main phases of the hexagonal structure of iron oxide (hematite) grown on the glass substrate. Besides, with increasing the annealing temperature from 400°C to 600°C, the crystal plates have been grown completely where the strongest peaks are related to (110) and (104) planes with 2Ө values of 35.83° and 33.22°, respectively. These major peaks are in agreement with the literatures [12-13, 15].

In order to investigate the phase formation percentage and also to identify the structural and microstructural parameters of the annealed samples, the Rietveld method with application of FullProf software is used as an efficient procedure. In this method the essential parameters such as the point group, the initial value of lattice parameters for hematite and the atoms position is given to the software as input data to obtain the diffraction pattern. This simulation is continued until the minimal difference between the obtained pattern and simulation pattern have been achieved. The result of this simulation is shown in figure 3.

In the X-ray diffraction pattern simulation, the assumption of broadening of the diffraction lines due to the non-spherical shape of the crystals and the high density of the crystalline nanomaterials defects is taken into account in the software, and the results are shown in table 1.

Table 1: The results of crystal size and strain at different annealing temperature for the $Fe_2O_3$ nanostructure.

| h | k | l | 2θ | α-$Fe_2O_3$ annealed at 400°C | |
|---|---|---|---|---|---|
|   |   |   |   | Size (nm) | Strain |
| 0 | 1 | 2 | 24.24 | 7.4 | 30.2 |
| 1 | 0 | 4 | 33.22 | 11.7 | 21.81 |
| 1 | 1 | 0 | 35.83 | 13.3 | 31.49 |
| 1 | 1 | 3 | 41.03 | 8.0 | 28.72 |
| 0 | 2 | 4 | 49.66 | 7.4 | 30.20 |
| 1 | 1 | 6 | 54.20 | 8.3 | 21.26 |
| 2 | 1 | 4 | 62.74 | 8.7 | 31.29 |
| 3 | 0 | 0 | 64.39 | 13.3 | 45.26 |

| h | k | l | 2θ | α-$Fe_2O_3$ annealed at 500°C | |
|---|---|---|---|---|---|
|   |   |   |   | Size (nm) | Strain |
| 0 | 1 | 2 | 24.20 | 14.0 | 13.93 |
| 1 | 0 | 4 | 33.17 | 20.2 | 14.78 |
| 1 | 1 | 0 | 35.76 | 13.4 | 4.31 |
| 1 | 1 | 3 | 40.96 | 13.6 | 13.90 |
| 0 | 2 | 4 | 49.57 | 14.0 | 13.93 |
| 1 | 1 | 6 | 54.11 | 14.8 | 14.81 |
| 2 | 1 | 4 | 62.61 | 13.8 | 13.57 |
| 3 | 0 | 0 | 64.25 | 12.8 | 11.44 |

| h | k | l | 2θ | α-$Fe_2O_3$ annealed at 600°C | |
|---|---|---|---|---|---|
|   |   |   |   | Size (nm) | Strain |
| 0 | 1 | 2 | 24.15 | 29.9 | 8.91 |
| 1 | 0 | 4 | 33.15 | 30.2 | 8.84 |
| 1 | 1 | 0 | 35.63 | 38.3 | 7.48 |
| 1 | 1 | 3 | 40.86 | 29.6 | 8.85 |
| 0 | 2 | 4 | 49.46 | 29.9 | 8.91 |
| 1 | 1 | 6 | 54.06 | 28.9 | 8.83 |
| 2 | 1 | 4 | 62.44 | 30.1 | 8.87 |
| 3 | 0 | 0 | 64.01 | 37.6 | 9.09 |

The average size of hematite crystals synthesized at annealed temperatures of 400°C, 500°C and 600°C are calculated to be 9.7nm, 14.5nm and 31.8nm, respectively. It can be observed that the average size of crystals had been increased with increasing the annealing temperature.

The lattice strain is affected by the lattice defects. If the lattice strain has significant changes in different reflections, it will cause anisotropic broadening of the diffraction lines. The mean values of lattice micro strain for all synthesized hematite phases at the annealing temperatures of 400, 500 and 600°C are 30.02, 12.58 and 8.72, respectively. As the annealing temperature increases, the average lattice strain will be decreased. In other words, the lattice defects are reduced. In the 400°C annealed sample, the lattice micro strain is slightly fluctuated for all reflections except for (104), (116) and (300) at around the mean value. For 500°C annealed sample, the micro strain lattice for all reflections except for (110) has slight fluctuations at around the mean value. For the 600°C annealed sample, the micro strain lattice for each reflection is only fluctuated around the average value. The anisotropic strain cannot be defined in any phases, and the crystals grow with the least amount of lattice defect.

The hematite nanostructured ultraviolet-visible absorption spectra at three annealing temperatures of 400°C, 500°C and 600°C are shown in figure 4a. The annealing temperature has affected the optical properties of the hematite nanostructure. At annealing temperature of 400°C and 500°C, the maximum absorption is in the range of 280 nm to 500 nm and then the absorbance decreased from 500 nm to 750 nm. However, for an annealing temperature of 600°C, the maximum adsorption range is observed from 280 nm to 570 nm and then it decreases. Therefore, by increasing the annealing temperature, the light absorption spectrum shifts to larger wavelengths.

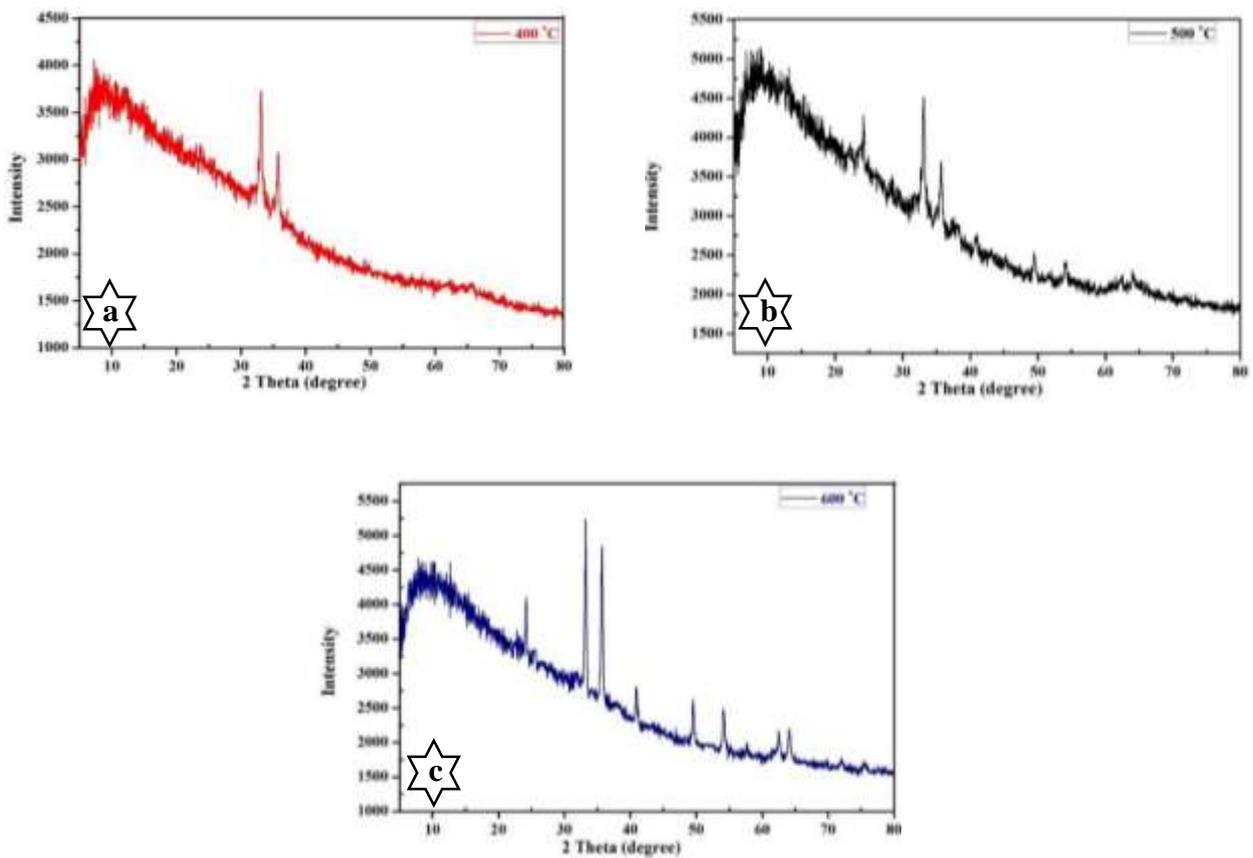

Figure 2 (Color online): The XRD patterns of annealing samples at a) 400°C, b) 500°C, and c) 600°C.

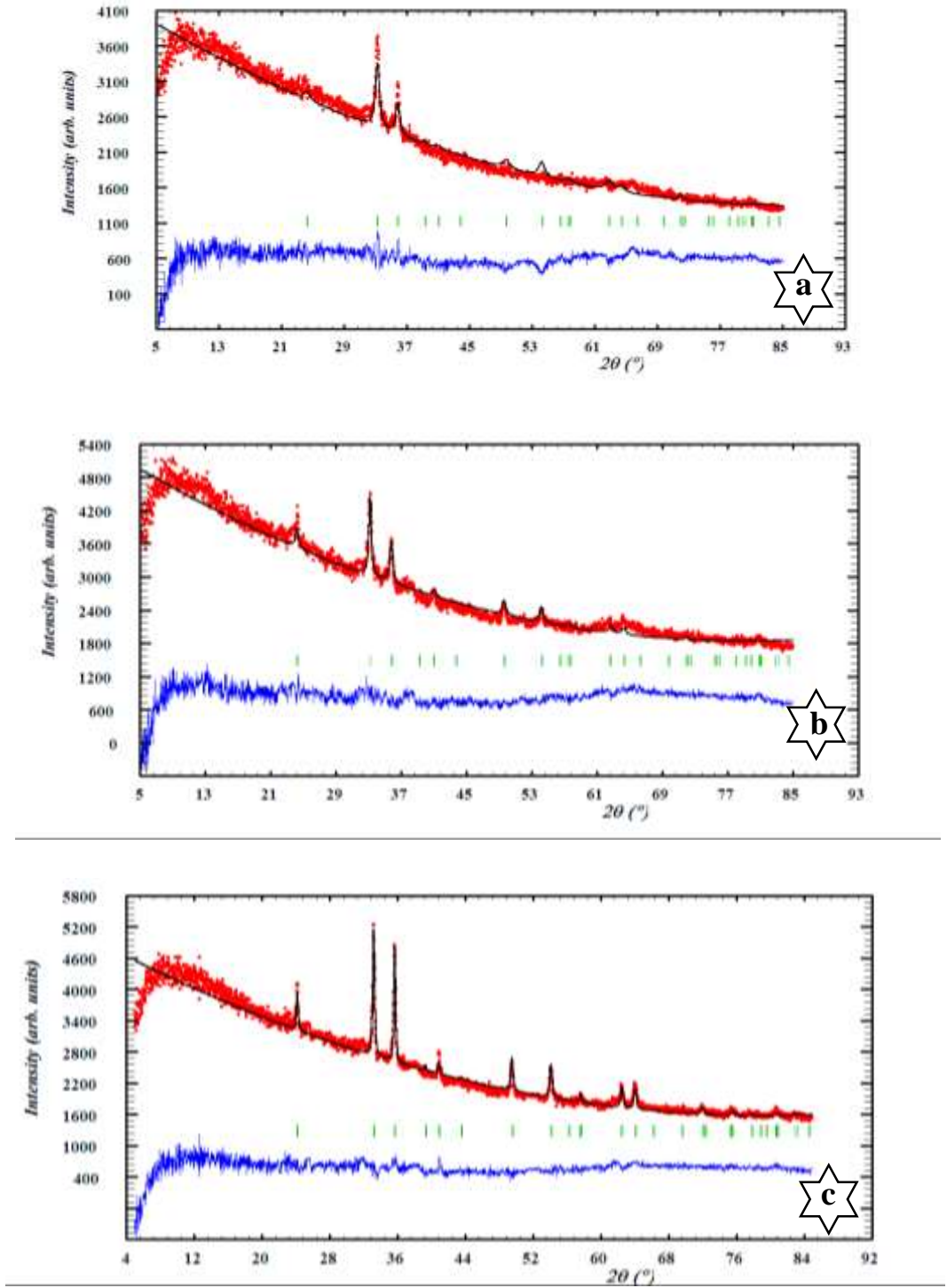

Figure 3 (Color online): The simulation diffraction pattern of iron oxide nanostructures at various annealing temperature on the glass substrate: a) 400°C, b) 500°C and c) 600°C.

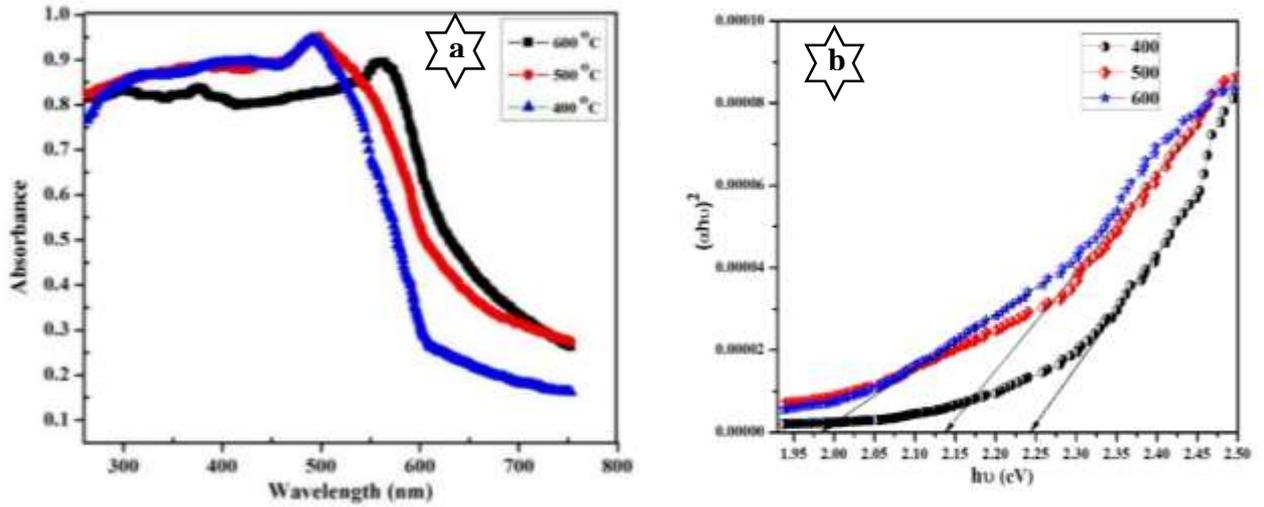

Figure 4 (Color online): a) The UV-VIS absorption for various annealed temperature from 400°C to 600°C. b) The plots of $(\alpha h\nu)^2$ versus $h\nu$ for annealed hematite at 400, 500 and 600°C.

The optical energy gap ($E_g$) can be calculated for the synthesized samples according to the following equation:

$$\alpha h\nu = A(h\nu - E_g)^n \qquad (3)$$

In this equation, α is the absorption coefficient, $h\nu$ is the photon energy, A represents a constant number, $E_g$ is the energy of band gap, and n is related to the indirect (n=2) band transition or direct band transition (n=0.5). The direct energy gap for hematite samples at annealing temperatures of 400, 500 and 600°C (Figure 4b) is obtained to be 2.24, 2.14 and 1.98 eV, respectively. The results showed that the energy gap is decreased by increasing the annealing temperature which can be attributed to the surface morphological features and structural properties.

### 3.2. Gas sensing mechanism

The sensing mechanism of iron oxide nanostructures (hematite) to ethanol gas can be described by the surface transfer of electrons from the oxygen agent to the adsorbed ethanol gas molecules (electron donors). As iron oxide (hematite) nanostructure is exposed to air, the oxygen molecules are absorbed on the surface of the iron oxide layer to form the $O^-$ and $O^{2-}$ spices. To do this, the electrons are taken from the conduction band of iron oxide. The kinetics reactions are shown in the form of the following relations:

$$O_2(gas) \rightarrow O_2(ads)$$
$$O_2(ads) + e^- \rightarrow O_2^-(ads)$$
$$O_2^-(ads) + e^- \rightarrow 2O^-(ads)$$

The molecular species of oxygen are active on the sample surface at temperatures below 150°C, whereas the adsorbents reacted with the gas molecules at temperatures between 150°C and 450°C.

The mechanism of ethanol detection using the iron oxide (hematite) sensor can be explained by two different oxidation pathways, the first one involves the ethanol oxidation through ethanol dehydrogenation, which forms the CH₃CHO mediator, and the second pathway involves the ethanol dehydration and the formation of C₂H₄. The choice of each of these two reactions is determined by the acid-base properties of the metal oxide surface. The dehydrogenation process base on oxide surfaces is much more likely, in addition to the fact that dehydrogenation processes usually occur at higher temperatures than dehydrated processes. Various researches have shown that ethanol is dehydrogenated at the temperature ranging from 100°C to 300°C to form acetaldehyde (CH₃CHO) as a mediator. By reacting acetaldehyde with oxygen anion which is adsorbed on the surface, the lattice electrons are transferred into the conduction band of α-Fe₂O₃ and thus the resistance is decreased and the acetate mediator (CH₃CO₂) decomposed into carbon dioxide and water. Ethanol gas reactions with hematite are presented in the following relations:

$$O_2(gas) \rightarrow O_2(ads)$$
$$O_2(ads) + 2e^- \rightarrow 2O^-(ads)(150 - 450°C)$$
$$C_2H_5OH + O^-_{ads} \rightarrow CH_3CHO_{ads} + H_2O + e^-$$
$$CH_3CHO_{(ads)} + 5O^-_{(ads)} \rightarrow 2CO_2 + 2H_2O + 5e^-$$

Briefly, the interaction of ethanol gas as a reducing agent with the chemically adsorbed oxygen on the sensor surface led to the release of the trapped electrons. Therefore, the density of charge carriers in the volume of the sensitive layer of iron oxide is increased. Increasing the density of electrons in the conduction band in the iron oxide layer resulted in reduction of the sensor resistance in the presence of reducing gas. After the ethanol gas exited the test chamber, air entered the chamber, and the above reverse mechanism occurred. Hence, the electrical conductivity returned to its original state. The gas sensor response depends on the surface chemical reactions of the oxygen species that lead to the adsorption and desorption of surface oxygen. The surface morphology in used material, the target gas sample, and operating temperature are the essential parameters that have an important influence on this reaction [56].

In order to measure the ethanol gas concentration, equation 4 is used due to the liquid phase equation.

$$C = \frac{22.4 \, \rho T V_s}{273 \, MV} * 1000 \qquad (4)$$

In this equation $C$ is the ethanol gas concentration in air in the term of ppm, $\rho$ is the density of desired liquid (g/mL), $T$ is the working temperature (K°), $V_s$ is the volume of the desired liquid (µL), $M$ is the molecular mass of the liquid (g/mol), and $V$ is the volume of the chamber (L). The density of ethanol is 0.789 g/mL and its molecular mass is 46.07 g/mol [57].

In this research, the measurable parameter of the sensor is the surface resistance of the synthesized flower-like hematite nanostructures. Hence, the percentage of sensitivity or the sensor response is calculated via the following equation:

$$Sensitivity \, (S_g)\% = \frac{R_{air} - R_{gas}}{R_{air}} * 100 \qquad (5)$$

where $R_{air}$ is the resistance of the sensor in the presence of air and $R_{gas}$ is the sensor resistance in the presence of ethanol. The response and recovery times are defined as the times required for a change in the resistance to reach 90% of the equilibrium value after the detected gas is injected and removed, respectively.

### 3.3. Effect of annealing temperature on the gas-sensing performance of the porous α-Fe$_2$O$_3$ nanostructures (hematite)

The annealing temperature had been affected the morphology of the hematite nanostructures samples. In order to detect ethanol gas, we have looked for a best iron oxides annealing temperature by synthesizing samples at three different temperatures of 400, 500 and 600°C. Three operating temperatures of 200, 250 and 300°C are investigated for a fixed ethanol concentration (200ppm) for each of above annealing temperature. In figure 5, the response curve of the iron oxide nanostructures at operating temperature of 200, 250 and 300°C is shown. For each operating temperature, the effect of annealing temperature (400, 500 and 600°C) is depicted. A summary of the percentage sensitivity of hematite samples in the presence of 200 ppm ethanol at different operating temperatures is also presented in table 2.

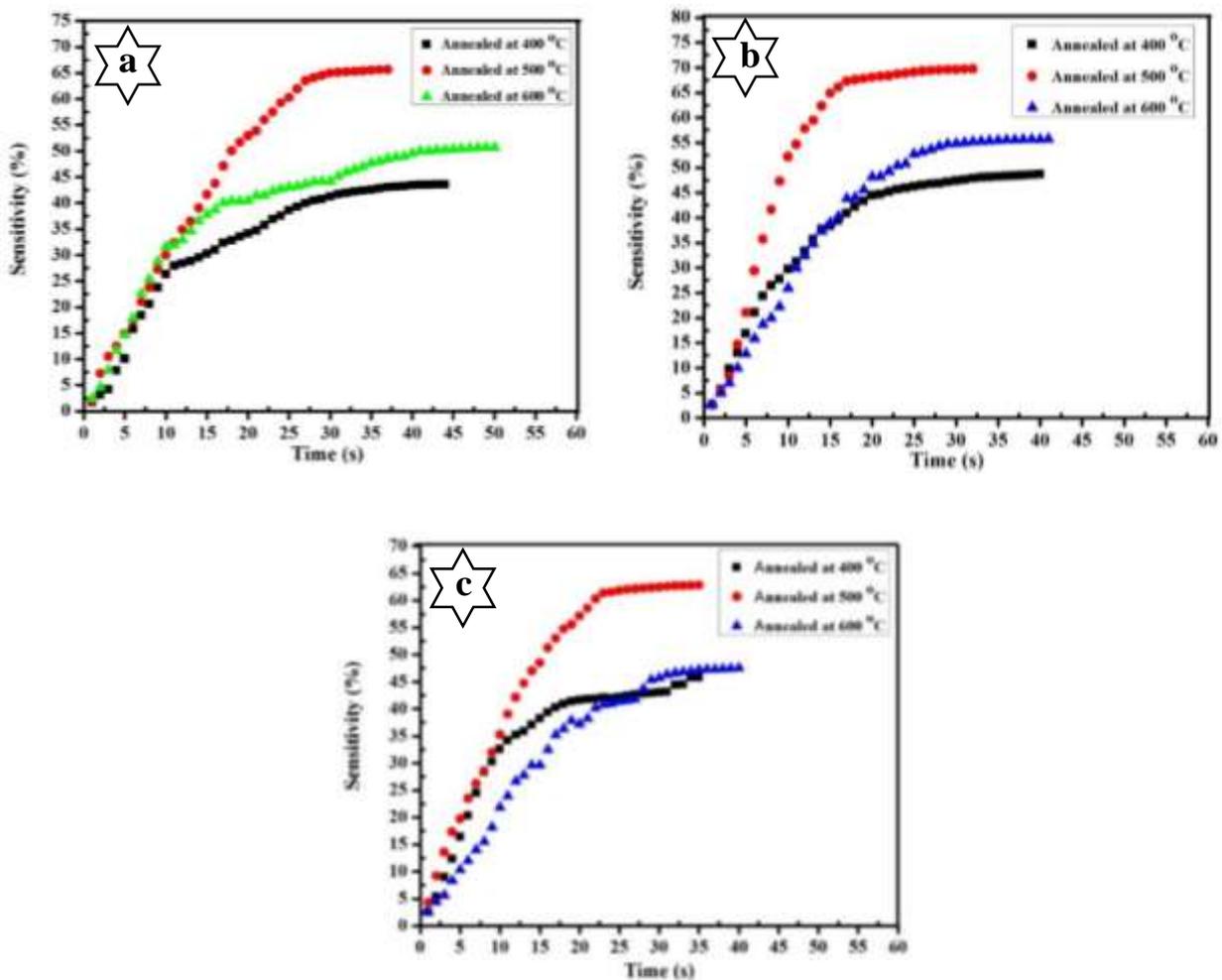

Figure 5 (Color online): The sensor response of the iron oxide (hematite) sampled to ethanol gas (200 ppm) which are annealed at 400, 500 and 600°C with the working temperature of a) 200°C, b) 250°C and c) 300°C.

Table 2: The percentage sensitivity of the synthesized samples for 200ppm ethanol concentration at various operating temperatures (°C).

| Sample | operating temperature | percentage sensitivity | Response time (s) |
|---|---|---|---|
| α-Fe$_2$O$_3$-400 | 200 | 43.57 | 39.6 |
|  | 250 | 48.70 | 36 |
|  | 300 | 45.81 | 32.4 |
| α-Fe$_2$O$_3$-500 | 200 | 65.61 | 33.3 |
|  | 250 | 69.75 | 29.7 |
|  | 300 | 62.82 | 31.5 |
| α-Fe$_2$O$_3$-600 | 200 | 50.67 | 45 |
|  | 250 | 55.68 | 36.9 |
|  | 300 | 47.49 | 36 |

According to the FESEM images (figure 1), the surface porosity of the hematite layer had been clearly changed for various annealing temperatures. The morphology had a great impact on the sensing behavior of ethanol gas. As can be observed from figure 5, the highest percentage of the ethanol gas sensor sensitivity is related to the iron oxide sample annealed at 500°C. The 500°C annealed sample with porous flower-like structures and high volume to surface ratio showed an extraordinary sensitivity and the oxidation-reduction interaction between ethanol gas and the adsorbed oxygen on the surface of hematite nanostructures. Therefore, the α-Fe$_2$O$_3$ sample annealed at 500°C has the best response to ethanol gas. In figure 6, the changes of the highest response of the synthesized hematite at different annealing temperatures according to the working temperatures of the sensor are shown. The highest response of the synthesized samples to ethanol gas belonged to the annealing temperature of 500°C.

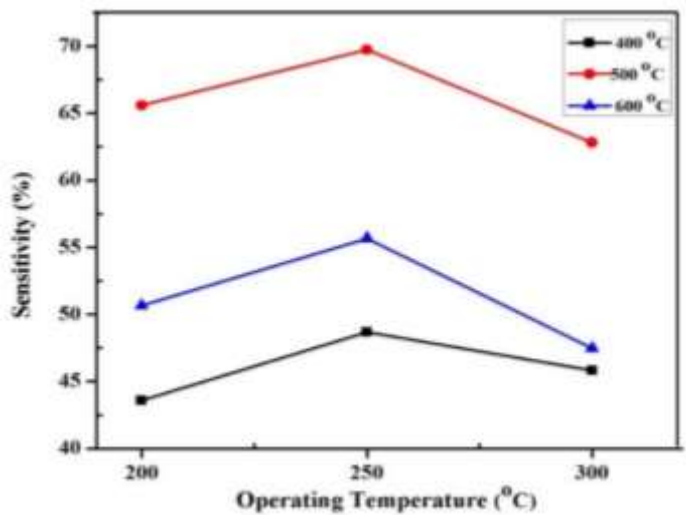

Figure 6 (Color online): The highest response of the synthesized hematite at various annealing temperatures in terms of operating temperatures.

### 3.4. Effect of operating temperature on the Gas-sensing performance of the porous α-Fe$_2$O$_3$ nanostructures (hematite)

In this section, we have looked for the best operating temperature for hematite sample which is annealed at 500°C (α-Fe$_2$O$_3$-500). In order to determine the optimum operating temperature of the nanostructured hematite sensor (α-Fe$_2$O$_3$-500), the sensing process is performed at different operating temperatures. The response curve of the sensor at operating temperatures of 150, 200, 225, 250, 300 and 350°C at the ethanol concentration of 200 ppm is presented in figure 7a. The maximum sensor response at different operating temperatures is also shown in figure 7b.

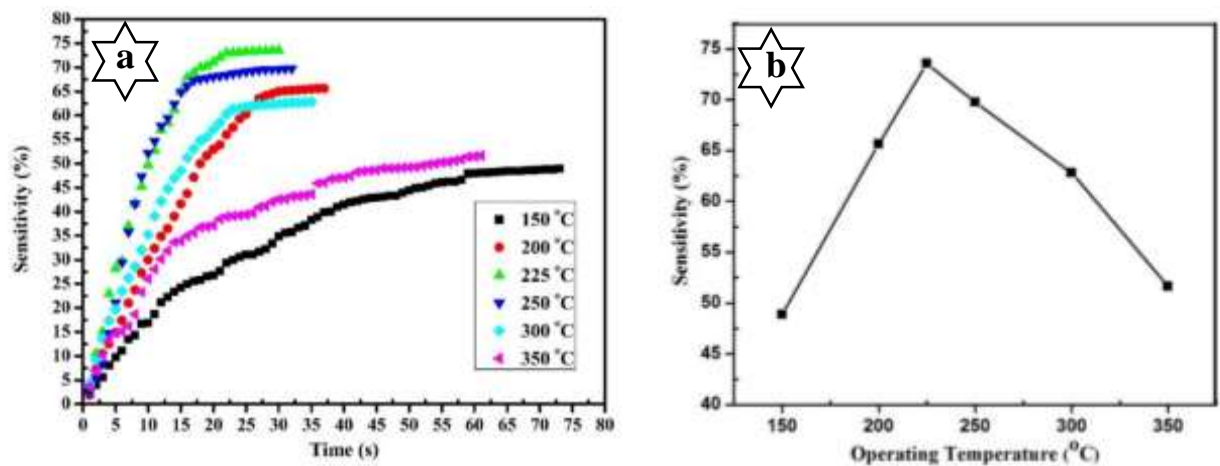

Figure 7 (Color online): a) The sensitivity curve of hematite flower-like hematite nanostructures sensor at operating temperatures of 150, 200, 225, 250, 300 and 350°C for the ethanol concentration of 200 ppm. b) The response of annealed at 500 °C versus the operating temperature of the ethanol sensor.

As shown in figure 7b, the response percentage of the hematite flower-like nanostructures against 200ppm ethanol at operating temperatures of 150, 200, 225, 250, 300 and 350°C are obtained at 48.88, 65.61, 73.55, 69.75, 62.82 and 51.63, respectively. The highest response is observed at the operating temperature of 225 °C. As the operating temperature raised above the optimum temperature (225 °C), the response rate of the sensors gradually decreased due to the insufficient attendance of the oxygen agents on the nanostructure surface to react with the ethanol gas molecules [58]. Besides, the sensitivity is decreased at operating temperatures below 150°C due to water absorption. In fact, at lower temperatures, the surface adsorption of water has gradually occurred. Due to the opposite effect of oxygen and moisture on conductivity, the rate of conductivity is decreased, which in turn reduces the surface activation energy. However, at higher temperatures, it can be observed that the activation energy is increased and the conductivity is changed significantly with the temperature.

In this temperature range, in addition to the moisture dissipation which is gradually increased with increasing temperature, the oxygen agent is absorbed as the O$^-$ species that are much more active than oxygen absorbed in the lower temperature range (O$^{2-}$ ions).

### 3.5. Effect of different ethanol concentration on the Gas-sensing performance of the porous α-Fe$_2$O$_3$ nanostructures (hematite)

The response of the iron oxide nanostructure sensor is obtained at different concentrations of ethanol gas from 50 to 1000 ppm which is represented in figure 8. The sensitivity of the sensors at different concentrations of ethanol gas is increased with increasing concentration from 50 ppm to 1000 ppm. The percentage of sensitivity at 225 °C for sensors at concentrations of 50, 100, 200, 600, 400, 800 and 1000 ppm are 52.40, 59.32, 73.55, 90.41, 93.04, 94.17, and 94.54, respectively. As it is shown in figure 8, in the presence of ethanol gas, the surface resistance of the flower-like nanostructure began to decrease due to the interaction of ethanol gas and the chemically adsorbed oxygen. After a while, the electrons are released and eventually the conductivity increased, so that the resistance difference in the presence and absence of ethanol is increased and resulting in improving the sensor response. Eventually, over time, these resistance changes became constant as can be observed in figure 9a. After the ethanol gas is released, the resistance increase, and the sensitivity of the sensor decrease. The resistance increased until it almost reached the surface resistance of the samples in the presence of air. It is worthwhile to mention that according to figure 9a, for a certain concentration of ethanol, the resistance of the hematite flower-like nanostructured returned to its original value, which represent the sensor reversibility during different sensing cycles. The sensitivity curve of hematite samples to different concentrations of ethanol at an optimum temperature of 225 °C is shown in figure 9b. At low concentrations of ethanol (less than 400 ppm) the ascending rate of the sensor response changes is almost linearly proportional to the increase of the ethanol concentration. However, as the ethanol concentration increases to more than 400 ppm, the rate of increase of sensor response changes slows down, and the sensor response becomes saturated.

However, as the ethanol concentration increases to more than 400 ppm, the rate of increase of sensor response changes slows down, and the sensor response becomes saturated. In fact, at lower concentrations, the sensor response is proportional to the gas concentration. However, at higher concentrations, due to the overall coverage of the surface by ethanol molecules and the inability of the sensor surface to absorb new ethanol molecules, the increasing rate of the response is decreased and the surface became saturated. As is shown in figure 9b, the limit of detection of ethanol gas concentration for the hematite nanostructure sensor is obtained at 50 ppm at the optimum temperature of 225 °C and the resolution of the sensor gas concentration is obtained at 50 ppm.

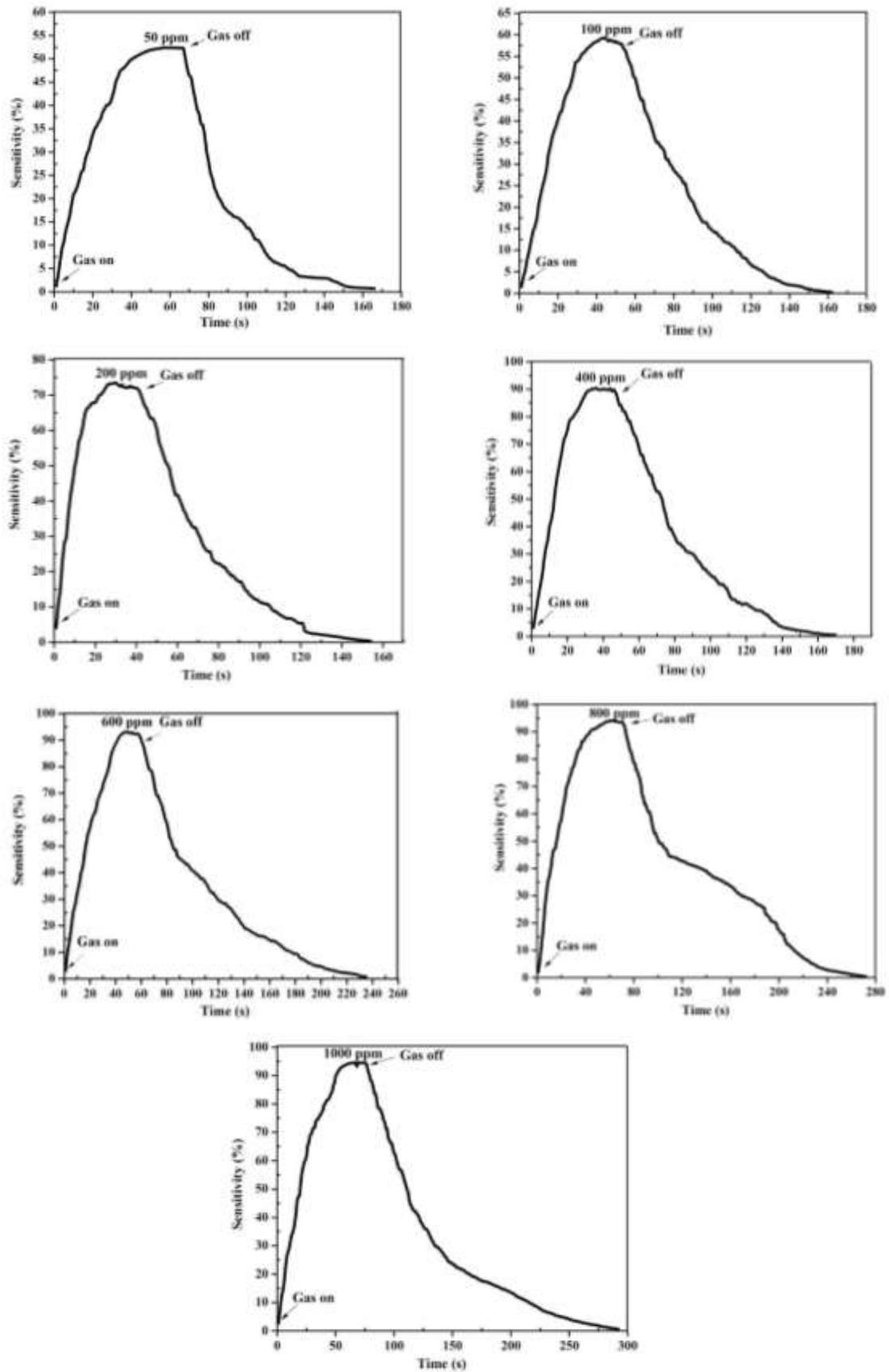

Figure 8: The sensitivity of the hematite flower-like nanostructure sensor from 50 ppm to 1000 ppm ethanol at operating temperature 225°C.

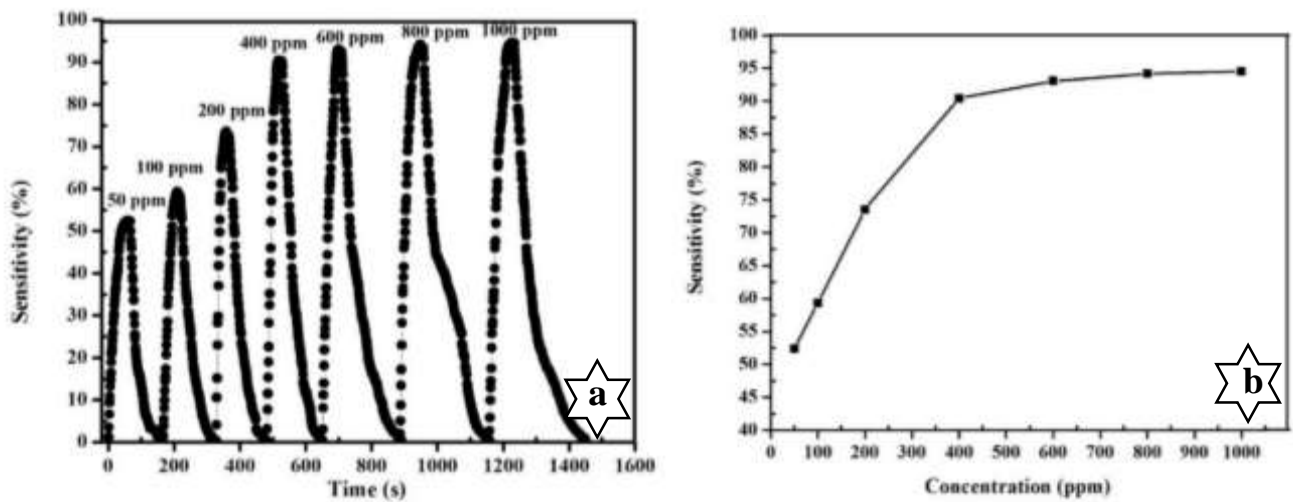

Figure 9: a) The sensitivity of the sensor to different concentrations of ethanol at operating temperature 225 °C. b) The sensitivity of hematite samples against to various concentrations of ethanol at optimum temperature (225°C)

### 3.6. The response time and recovery time of the porous α-Fe$_2$O$_3$ nanostructures (hematite)

The response time is defined as the time required for the resistance in the presence of gas (Rgas) to reach 90% of its final value and the recovery time is the time required for the air resistance (Rair) to reach 90% of its final value [59]. The gas response is rapid at first and then reach to a constant value. The response time and the recovery time of the iron oxide nanostructure sensor (hematite) at the optimum temperature of 225°C at different ethanol concentrations are presented in the table 3.

Table 3: The response time and recovery time of hematite nanostructure sensor at 225°C at different concentrations of ethanol.

| Concentration (ppm) | 50 | 100 | 200 | 400 | 600 | 800 | 1000 |
|---|---|---|---|---|---|---|---|
| **Response time (s)** | 51.3 | 39.6 | 27.0 | 32.4 | 43.2 | 55.8 | 58.5 |
| **Recovery time (s)** | 83.1 | 97.2 | 102.6 | 110.7 | 159.3 | 180.9 | 194.4 |

The lowest response time belonged to an ethanol concentration of 200 ppm (about 27 seconds) and the highest response time belonged to an ethanol concentration of 1000 ppm (about 58.5 seconds). Besides, the shortest recovery time for 50 ppm ethanol concentration is 89.1 seconds. The response time trend has decreased for the ethanol concentration from 50 ppm to 200 ppm, but for the ethanol concentration from 200 to 1000 ppm, the response time had been increased. It is worthwhile to mention that with increasing the ethanol concentration, the recovery time has been increased.

### 4. Conclusion

In summary, the Fe$_2$O$_3$-500 nanostructure has been successively fabricated by two facile steps of the hydrothermal and annealing method. The XRD and FESEM studies indicated the flower-like structure of hematite with a flower leaf average diameter of 100 nm. The band gap value has been also obtained from UV-VIS absorption spectra. Our data obviously reveal the enhancement in effective surface area for the fabricated sample. Moreover, the α- Fe$_2$O$_3$-500 nanostructure exhibited high response and fast response recovery time to ethanol gas exposed from 50 ppm to 1000 ppm. Our results clearly demonstrate that the proposed nanostructure is a valuable tool for gas-sensing applications.